\documentstyle[12pt, epsf, rotate]{article}
\title{RX J0720.4--3125 as a Possible Example of the  Magnetic 
Field Decay of Neutron Stars}
\author{D. Yu. Konenkov$^1)$ and S. B. Popov$^2)$ \\
        1) A.F.Ioffe Institute of Physics and Technology \\
        194021 St.Petersburg, Russia, \\
        {\it e-mail: dyk@astro.ioffe.rssi.ru} \\
        2) Moscow State University, 119899, Moscow, Russia \\
        {\it e-mail: polar@xray.sai.msu.su}
       }    
\begin{document}
\maketitle

\begin{abstract} 

We studied possible evolution of the spin period and the magnetic
field of the X-ray source RX J0720.4-3125 assuming this source to be an 
isolated neutron star accreting interstellar medium. Magnetic field of the
source is estimated to be $10^6 - 10^9$ G, and it is difficult to explain
observed spin period  8.38 s without invoking hypothesis of the 
magnetic field decay. We used the model of ohmic decay of the crustal 
magnetic field. The estimates of accretion rate ($10^{-14} - 10^{-16} 
M_\odot/{\rm yr}$), velocity of the source relative to interstellar
medium ($10 - 50 $ km/s), neutron star age
($2\cdot 10^9 - 10^{10}$ yrs) are obtained.

\end{abstract}

\begin{center}

Received December 20, 1996; in final form, February 2, 1997 \\
Published: August 1997 Vol.23 p. 569\\
 (Pisma v Astronomicheskii Zhurnal, Astronomy Letters: \\ 
http://kisa.iki.rssi.ru/~pazh/)

\end{center}

\section{Introduction} 

Isolated neutron stars (INS) have received special attention in the last few 
years. The idea of observing such objects in the X-ray range has 
emerged rather long ago (Ostriker et al., 1970). Treves and Colpi (1991) 
supposed that INSs accreting matter of the interstellar medium (ISM) can be 
observed with the $ROSAT$ satellite in the UV and X-ray ranges. 

The estimations of energy characteristics, spectra and possibilities of 
observing INSs were made in several works (see, for example, B\"ohringer 
et al., 1987; Treves and Colpi, 1991; Blaes and Madau, 1993). A large 
number of works is devoted to the study of the spatial distribution of INSs 
(see, for example, Blaes and Rajagopal, 1991). 

In the mid-1996, Haberl et.al. reported the discovery of the pulsating 
source RX J0720.4--3125 with the $ROSAT$ satellite in the soft X-ray 
range. We shall use two observational characteristics of this source (Haberl 
et.al., 1996): the period $p = 8.38$ s and the blackbody temperature 
$T = (79 \pm 4)$ eV. 

Using the hypothesis that this source is an INS accreting matter from the 
ISM (we do not follow alternative hypotheses: further observations will show 
if it is true or not), we estimate the accretion rate and the magnetic field
strength of this source. 
We show that the INS could not increase its period to the observed value 
over the Universe lifetime if we assume that it was born with the present-day 
magnetic field strength. Finally we conclude that the 
magnetic field of this neutron star (NS) had to decay. Then, 
using the model of ohmic dissipation of the magnetic field we calculate the 
magnetorotational evolution of the NS. 

The paper is organized as follows. In Section 2, we present the reference 
material on the evolution of the neutron star period and on the model of 
ohmic dissipation of the magnetic field; in Section 3, the analytical
estimates 
of the neutron star parameters are made; the evolutionary tracks of the INS 
are computed in Section 4; and the conclusions are given in Section 5.

\section{Evolution of the spin period and the magnetic field of the 
neutron star} 

\subsection{Spin period}

In the low-density plasma an INS may have four possible evolutionary states 
(Lipunov, 1987):  ejector, propeller, accretor, and georotator. A
particular 
state is determined by the relation between four characteristic radii: the
light 
cylinder radius $R_l=c/\omega $; the stopping radius  $R_{st} $; 
the radius of the gravitational 
capture $R_G=(2GM)/v_\infty ^2$; 
and the corotation radius  $R_{co}=(GM/\omega ^2)^{1/3}$. Here, 
$M$ is the NS mass, $c$ is the speed of light, $\omega$ is the spin
frequency, 
and $v_{\infty}$ is the NS velocity with respect to the ISM. 

The relationship between these radii determines two critical periods: $P_E$ 
and $P_A$ which separate different stages of the NS evolution. These 
periods can be estimated using the formulae (Lipunov, 1987):

\begin{equation}
P_E= 2\pi \left(\frac {2\, k_t}{c^4}\right)^{1/4}
\left(\frac {\mu ^2}{v_{\infty}\dot{M}}\right)^{1/4},
\quad R_l < R_G,\\
\end{equation}

\begin{equation} P_A= 2^{5/14}\pi (GM)^{-5/7}
\left(\frac {\mu ^2}{\dot{M}}\right)^{3/7},  
\quad R_A<R_G.\\
\end{equation}  
Here $\mu$ is the magnetic dipole moment, 
$\dot M \equiv \pi R_G^2 \rho v_{\infty}$ is the accretion rate, 
$\rho$ is the ISM density, and $k_t$ is a dimensionless constant of the 
order of unity. 

If $p < P_E$, then NS is at the ejector stage; if $P_E < p <
P_A$, we have the NS at the propeller stage; and if $p > P_A$
and $R_{st} < R_G$, then NS is an accretor. In some cases the
situation is possible when $p > P_A$, but $R_{st} > R_G$ and
accretion is impossible because of the formation of the geolike
magnetosphere. However, we shall not be interested in the
georotator stage since we consider the accreting NSs for which
$R_{st} < R_G$.

At the ejector stage, the evolution of the spin period is
determined by the losses of the INS kinetic energy due to
magnetic dipole radiation:

\begin{equation}  
\dot p = \frac{8\pi^2 R^6}{3c^3 I} \cdot   \frac{B^2(t)} {p},
\end{equation}
where $R$ is the NS radius, $I$ is the moment of inertia, and $B
=\mu/R^3$ is the magnetic field strength. 

At the propeller stage, the NS spin-down rate is determined by
the transfer of the angular momentum to the surrounding matter
(Illarionov and Syunyaev, 1975):

\begin{equation}
\dot p = \frac{2^{2/7}}{\pi}\frac{(GMR^2)^{3/7}}{I}
p^2 B^{2/7} \dot {M}^{6/7}.
\end{equation}

At the accretion stage, the NS is acted upon by two moments of forces:

\begin{equation}
\frac{d(2\pi I/p)}{dt} = K_{sd} + K_{turb},
\end{equation}
$$ K_{sd}=-k_t \frac{\mu^2}{R_{co}^3}.
$$ 
Here $K_{sd}$ is the braking  moment of forces which can arise due to the 
possible turbulization of the ISM. $K_{turb}$ acts randomly and can 
either spin up or spin down the NS (Lipunov and Popov, 1995). 

The change in the period of an accreting INS is due to its
interaction with the turbulized ISM. This introduces specific
features into the problem of the period evolution. If we adopt
the hypothesis of spin acceleration of the NS in the turbulized
ISM (Lipunov and Popov, 1995), the new characteristic period
emerges,

\begin{equation}
P_{eq}=960 k_t^{1/3}\mu_{30}^{2/3}I_{45}^{1/3}\rho_{-24}^{-2/3}
v_{\infty_6}^{13/3}v_{t_6}^{-2/3} M_{1.4}^{-8/3}\,\, {\rm sec}=  
\end{equation}

$$ =3450
k_t^{1/3} \mu_{30}^{2/3} I_{45}^{1/3} \dot M_{-15}^{-2/3}
v_{\infty_6}^{7/3} v_{t_6}^{-2/3} M_{1.4}^{-4/3}\,\, {\rm sec},
$$
where $\mu_{30}$ is the magnetic dipole moment in units of $10^{30}$ Gs
cm$^2$, $I_{45}$ is the moment of inertia in units of $10^{45}$ g cm$^2$, 
 $\rho_{-24}$ is the ISM density in units of $10^{-24}$ g cm$^{-3}$, 
$v_{\infty_6}$ is the 
NS velocity relative to the ISM in units $10^6$ cm/s, and $v_{t_6}$ 
is the turbulent velocity in units of
$10^6$ cm/s. The period $P_{eq}$ corresponds to the NS rms
rotation rate obtained from the solution of the corresponding
Fokker--Planck equation. In reality, the rotational period of INS
fluctuates around this value. Note that here we make a more
accurate estimate of the period than that made in the work of
Lipunov and Popov (1995) (we are grateful to M.E. Prokhorov for
his assistance in performing the corresponding calculations). We
take into account the three-dimensional character of turbulence,
i.e. the fact that the vortex can be oriented not only in the
equatorial plane but also at any angle to this plane. In this
case, diffusion occurs in the three-dimensional space of angular
velocities.

\subsection{Ohmic dissipation of the crustal magnetic field
of the neutron star}

The magnetic field decay in the neutron star crust has been investigated by 
many authors (see, f.e., Urpin and Muslimov, 1992). Such a decay is 
governed by the induction equation

\begin{equation}
\frac{\partial {\bf B}}{\partial t}=
-\frac {c^2}{4 \pi} \nabla \times \left( \frac{1}{\sigma}\nabla
\times {\bf B} \right)
+
\nabla \times \left( {\bf v} \times {\bf B}\right),
\end{equation}
where $\sigma$ is the conductivity; ${\bf v}$ is the velocity of
the crust motion; and ${\bf v} = 0$ at the ejector and propeller stages.
At the accretor stage in spherically symmetrical case, 
$ {\bf v} = (-v_r, 0, 0)$, with the component 

$$v_r=\frac{\dot M}{4 \pi r^2 \rho(z)},
$$ 
where $r$ is the distance to the NS center, $\rho(z)$ is the
density of matter at the depth $z = R-r$. Equation (7) is highly
simplified in the case of the dipole field and reduced to the
one-dimensional parabolic equation for the vector potential. The
boundary conditions are set in the same ways as in the work of
Urpin and Muslimov (1992).

The conductivity $\sigma$ is determined basically by the scattering of 
electrons by phonons and impurities: 

$$
\frac{1}{\sigma}=\frac{1}{\sigma_{ph}}+\frac{1}{\sigma_{imp}}.  
$$

The phonon conductivity $\sigma_{ph}$ depending on density and 
temperature dominates at high temperatures and at moderately high densities. 
At lower temperatures and higher densities the impurity conductivity 
$\sigma_{imp}$ prevails. The impurity conductivity is independent of the 
temperature but is the function of concentration and the charge of impurities 
which are characterized by the parameter 

$$Q= \frac{1}{n} \sum_{n'} n' \left( Z-Z' \right)^2, 
$$ 
where $n$ and $Z$ are, respectively, the concentration and the
charge of the major ion species, and $n'$ and $Z'$ are the
concentration and the charge of the impurity, the summation is
over all impurity species. The analytical formula for the
impurity conductivity was obtained by Yakovlev and Urpin (1980),
while the phonon conductivity was taken from the work of Itoh
et al. (1993). The initial field is assumed to be concentrated
mainly in the surface layer of a certain thickness. The density
of matter, $\rho_0$, at the inner boundary of this layer is the
parameter of the model. The quantity $Q$ is assumed to be
independent of the depth.

The main results of the calculation of the decay of the INS
magnetic field are the following. The magnetic field dissipation
proves to be intimately related to the NS thermal evolution. For
the {\it standard} cooling, at which the NS neutrino luminosity is
determined chiefly by the modified URCA-processes (Pethick,
1992), in the first million years the field decays by a factor
of $2 - 1000$, depending on the initial depth of the current layer
and on the adopted equation of state for the star core (Urpin
and Konenkov, 1997). As the NS cools down, the conductivity
increases and the field decay slows down. The decay rate at the
later stage depends on $\sigma_{imp}$ and, therefore, on $Q$. For
example, when $Q = 0.01$, the field does not decrease
in the subsequent $10^8$ yrs. However, as soon as the magnetic
field diffuses through the entire crust and reaches the
superfluid core (after $ \sim 10^9$ yrs at $Q = 0.01$), the
decay becomes exponential.

\begin{figure}
\epsfxsize=12cm  
\centerline{{\epsfbox{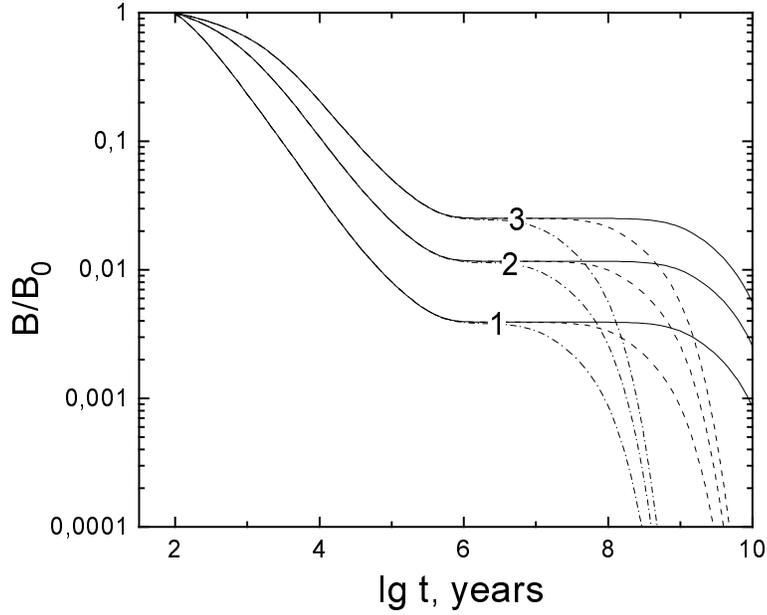}}}
\caption{ 
The change of the surface magnetic field of the isolated neutron star 
with time for the model of standard cooling. Curves 1, 2, and 3 correspond 
to the initial depths of the current layer, $10^{11}$, $10^{12}$, and
$10^{13}\,$ g cm$^{-3}$, respectively. The solid curves correspond to 
Q = 0.001; the dashed curves, to Q = 0.01; the dot-dashed curves, to Q = 0.1. 
}
\end{figure}

Figure 1 shows the decrease of the NS surface magnetic field
with time for various parameters $\rho_0$ and $Q$. We perfomed
our calculations for the model of the NS based on the
Friedman--Pandharipande (1981) moderately hard equation of state
in the star core, with the mass of the neutron star $M = 1.4M_{\odot}$, 
the radius $R = 10.6\,$km, and the crust thickness $\Delta R =
940\,$m (Van Riper, 1988).  It is apparent that the decay at the
initial stage ($t < 10^6\,$yr) is determined by the initial depth
of the current layer and that the subsequent decay rate depends
on the amount of impurities.

Accretion may affect the field evolution. First, it heats the NS
crust (Zdunik et al., 1992) and, therefore, decreases the
conductivity.  Second, the flux of matter toward the star center
arises; it tends to bring the field to deeper layers.
However, calculations show (Urpin et al., 1996) that accretion with the
rate $\dot M < 10^{-14}M_{\odot}\,$yr$^{-1}$ speeds up
insignificantly the field decay.

\section{ Estimates of parameters of RX J0720.4--3125}

Using the hypothesis that we observe an accreting isolated  NS and taking the
observed period $p = 8.38$ s and the temperature $T = (79 \pm
4)$ eV (Haberl et. al., 1996), we can obtain the constraints
on the magnetic field, accretion rate, and luminosity. 

1. If we really observe the accretor, then $p > P_A(B, \dot M)$, where 
$P_A$ is defined by (2). This yields the first constraint on $B$
and $\dot M$: 

$$
B < 4.6 \cdot 10^9 M_{1.4}^{5/6} R_6^{-3} \dot M_{-15}^{1/2}\,
                  {\rm G}, \eqno(8) 
$$

2. We see the pulsating radiation, it means that the accreting matter is 
funneled by the magnetic field onto the polar caps. Using the known relation 
$R_{\rm cap}= \sqrt{(R/R_A)} \cdot R,$ connecting the polar cap
radius $R_{cap}$, the Alfven radius $R_A$, and the NS radius, we
can estimate $R_{cap}$ as 

$$
R_{\rm cap}=0.53
  \cdot B_9^{-2/7} \dot M_{-15}^{1/7} R_6^{9/14}M_{1.4}^{1/14}\,{\rm km,}
$$
where $B_9 = B/10^9$ G. From the condition $R_{cap} < R $  (or 
$R_A > R$) we obtain another constraint: 

\begin{equation}
B>3.4
\cdot 10^4 M_{1.4}^{1/4}R_6^{-5/4} \dot M_{-15}^{1/2}\,{\rm G}.
\end{equation}

3. And finally, we know the polar cap temperature. From the condition 

$$ 2 \pi R_{\rm cap}^2 \sigma T^4 = \frac{GM\dot M}{R}
$$ 
we find the additional relation between $B$ and $\dot M$: 

\begin{equation}
B=2.5\cdot 10^7 \left(T/79 \,\, {\rm eV}\right)^7 R_6^4 M_{1.4}^{-3/2} 
                             \dot M_{-15}^{-5/4}\,{\rm G}.
\end{equation}
Combining inequalities (8) and (9) and condition (10) yields the allowed 
range of the $B$ and $\dot M$ values: 
\begin{eqnarray}
& 2\cdot 10^5< B[{\rm G}] < 10^9,\nonumber \\
& 6\cdot 10^{-17}< \dot M [M_\odot /{\rm yr}] < 4 \cdot 10^{-14}.
\end{eqnarray}
This range of possible $\dot M$ values corresponds to the allowed range of
luminosities $L$: 

$$ 7 \cdot 10^{29} < L \rm{ [erg/s]} < 5 \cdot 10^{32}. 
$$

Let us estimate the spin-down time to $p = 8.38$ s assuming that the NS was 
born as a radio pulsar with $p_0 \ll p_E$. The time of spin-down
to $p= p_E$ at a constant magnetic field is deduced from (3) and (1): 

$$ t_E=\frac{3 c^3 I} {16 \pi^2 R^6 B^2} P_E^2=
$$
$$ = 4\cdot 10^{12}
\dot M_{-15}^{-1/2} I_{45} v_{\infty_6}^{-1/2} R_6^{-3}B_{9}^{-1} 
                                        \,{\rm yrs}.
$$
For $B < 10^{10}$ G and $\dot M < 10^{-13} M_{\odot}\,$yr$^{-1}$
the spin-down time exceeds the age of the Universe. The
spin-down at the propeller stage takes an additional time. Thus,
we come to the conclusion that with the estimated values of the
magnetic field strength, the accretion rate, and the velocity of
motion relative to the ISM the NS could not spin down to $p =
8.38$ s. This means that in the past the given INS had a larger
magnetic field that has decayed over the time of evolution. 

\section{The evolutionary tracks of neutron star on the 
$B-P$ diagram }

\begin{figure}
\epsfxsize=12cm  
\centerline{{\epsfbox{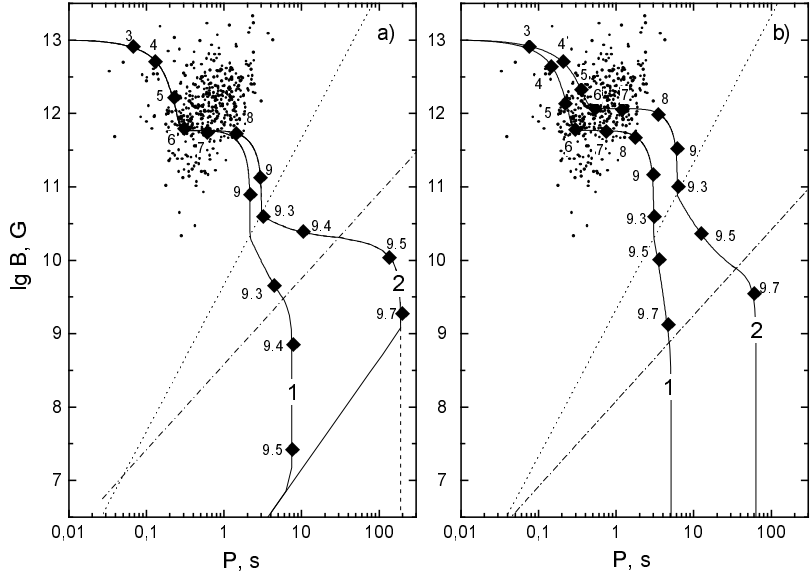}}}
\caption{ 
 The evolutionary tracks of the neutron star for the accretion rates
$\dot M = 
10^{-15} M_{\odot}\,$yr$^{-1}$ (a) and $\dot M = 10^{-16} M_{\odot}\,$
yr$^{-1}$ (b). The model 
parameters are described in the text. The dashed lines correspond to $p = 
P_E$; the dot-dashed lines, to $p = P_A$. The dashed line in Fig. 2a 
shows for the second track the neutron star evolution with no acceleration in 
the turbulized intestellar medium. The numbers near the marks in tracks 
denote the logarithm of the neutron star age in years. The observed radio 
pulsars are indicated by dots (Taylor et al., 1993). 
}
\end{figure}

The evolution of the spin period was calculated from
formulae (3) and (4) taking into account the decay of the
magnetic field at the ejector and propeller stages.  Since the
treatment of the spin-up rate in the turbulized ISM at the
accretor stage is not quite a simple matter, we applied to the
description of the period evolution the following simplifying
model. When the NS gets to the accretor stage, the spin-down
momentum substantially exceeds the spin-up momentum because the
acceleration occurs in the turbulized ISM and there is no
constant spin-up momentum, unlike in a binary system. However, we
can obtain an analog of the equilibrium period (Section 2.1)
which corresponds to the stationary solution of the
Fokker--Planck equation (Lipunov, 1987; Lipunov and Popov,
1995). Therefore the evolution of the rotational period
at this stage was treated as a stationary spindown from
$P_A$ to $P_{eq}$ after which the period was set to be $P_{eq}$,
which, in its turn, changed due to the magnetic field decay.

We calculated the magnetic and spin evolution of the NS
with mass $M = 1.4 M_{\odot}$ for the accretion rates $10^{-15}
M_{\odot}$yr$^{-1}$ and $10^{-16} M_{\odot}$yr$^{-1}$,
moving through the ISM with density $\rho = 10^{-24}$ g/cm$^3$.
Such accretion rates at the given ISM density correspond to the
velocities of motion $19$ and $41$ km/s. We assumed the NS to be
born as an ordinary pulsar with a short period ($p_0 = 0.01$ s)
and a "standard" initial magnetic field $B_0 = 10^{13}$ G. The
model of the ohmic decay of magnetic field allows one to obtain
both the high and low dissipation rate. However, in this case we
should fit the parameters of the model so as to obtain the
accreting INS with a period of 8.38 s and a field of $2 \cdot
10^7$ G for the accretion rate $\dot M = 10^{-15} M_{\odot}$
yr$^{-1}$ or with a field of $4 \cdot 10^8$ G for the accretion
rate $10^{-16} M_{\odot}$ yr$^{-1}$ (10). Therefore, the field
should decay with a moderate rate to give the NS enough time to
slow down to $p = 8.38$ s. Moreover, it is desirable at the
initial stage of evolution to provide the agreement with
magnetic fields and periods of the observed radio pulsars. Basing
on these assumptions we can obtain the estimates of the
parameters $\rho_0$ and $Q$, as well as the lower limit on the
INS age.

The evolutionary tracks of the NS for the accretion rates 
$\dot M = 10^{-15} M_{\odot}$ yr$^{-1}$ and 
$\dot M = 10^{-16} M_{\odot}$ yr$^{-1}$ are shown
in panels a) and b) of the Fig.2. 

Tracks 1 in both panels of Fig. 2 illustrate the evolution with
the maximum possible rate of magnetic field dissipation with
time. The model parameters for the accretion rate $\dot M =
10^{-15} M_{\odot}$ yr$^{-1}$ are: $\rho_0 = 3 \cdot 10^{13}$
g/cm$^3$, $Q = 0.02$, and $v = 10^6$ cm/s. The NS was born at the
point ($p_0, B_0$) at the left upper corner of the $B-P$
diagram. In the first $10_6$ yrs, the field decays by about 20
times, while the NS slows down to $p \approx 0.3$ s. At that
time the NS represents a typical radio pulsar.  Subsequently,
due to the cooling, the field decay slows down. In the next
$8\cdot 10^7$ yrs the field decays to $4.5 \cdot 10^{11}$ G, the
period increases, and the radiopulsar dies.  However, the
ejector stage continues for $1.2 \cdot 10^9$ years.  During this
stage, the field decays to $2.5 \cdot 10^{10}$ G, the period increases to
$2.1$ s, and the NS goes over to the propeller stage (1) which
lasts $\sim 10^9$ yrs. At this stage, further deceleration
of the NS occurs, according to (4), to $p = 5.7$ s, and the field
decays to $3 \cdot 10^9$ G. The star goes over to the accretor stage
and becomes the source of periodic X-ray radiation. The core
heating speeds up insignificantly the field decay (Urpin et.al.,
1996). The field decays to $2 \cdot 10^7$ G over $4 \cdot 10^9$ yrs.
The period does not increase because the field is weak enough.
As soon as $P_{eq}$ becomes equal to the current period, the
spin-up of the NS due to the interaction with the ISM may
occur. The period fluctuates around $P_{eq}$, but we do not
estimate here the amplitude of these fluctuations.

Tracks 2 illustrate the evolution with slower field decay. Such
a decay can be obtained, for example, for the crust with lower
impurity content. For $\dot M = 10^{-15} M_{\odot}$ yr$^{-1}$,
we set $Q=0.01$, while the other parameters remained the same as
for track 1 in Fig. 2a. Tracks 1 and 2 coincide at the initial
stage of evolution, when the dissipation rate does not depend on
the impurity concentration. However, in $10^7$ yrs the tracks
diverge. As a result, the star resides at the ejector stage $2
\cdot 10^9$ yrs going over to the propeller stage with $p = 3$
s and $B =4 \cdot 10^{10}$ G. The spin-down rate at the
propeller stage is much higher than in the first case for three
reasons: due to the longer spin period, the stronger
magnetic field during the NS transition from the ejector to the
propeller stage, and to the lower rate of the field decay caused by the
lower value of the $Q$ parameter. At the accretor stage, the NS
spin period decreases to 190 s over $2 \cdot 10^9$ yrs, and when
the field decays to $4 \cdot 10^8$ G, the effect of acceleration in
the turbulized ISM may begin to act. The track in the absence of
such acceleration ($v_t = 0$) is shown by the dashed line. In this
case the final period turns out to be about 200 s. The turbulent
acceleration may decrease this period. However, the
self-consistent calculation of the evolution of the NS period at
the accretor stage with allowance for the magnetic field decay
would require the solution of the Fokker--Planck equation for
the distribution function in the space of angular velocities,
which is beyond the scope of this work.

At the accretion rate $\dot M = 10^{-16} M_{\odot}$ yr$^{-1}$ we
used for track 1 the following parameters: $\rho_0 = 3 \cdot
10^{13}$ g/cm$^3$, $Q = 0.01$, and $v_t = 10^6$ cm/s. The NS
comes to the propeller stage in $2.7
\cdot 10^9$ yrs having a period of 3.1 s and a magnetic field of 
$2 \cdot 10^{10}$ G.  The transition to the accretor stage
occurs when the NS has a period of 4.9 s and a magnetic field of
$7 \cdot 10^8$ G. Deceleration at this stage does not occur because
the magnetic field is low. The full time taken for the field decay to
$4 \cdot 10^8$ G is $6 \cdot 10^9$ yrs. In this case $p = 5$ s. The second
track in Fig. 2b was calculated for the greater initial depth of
the current layer corresponding to $\rho_0 = 6 \cdot 10^{13}$
g/cm$^3$. As a consequence, at the initial stage of evolution the
magnetic field decays with a slower rate, the ejector stage
lasts $2.1 \cdot 10^9$ yrs, and the propeller stage lasts $1.9 \cdot
10^9$ yrs. For the field $4 \cdot 10^8$ G, the period is 63 s.

\section{Conclusion} 

The observed period and the temperature of the X-ray source RX
J0720.4-- 3125 can be accounted for by using the hypothesis for
the ISM accretion onto an old INS. We showed that the NS
magnetic field is low in this case ($B < 10^9 $ G). The time
taken for deceleration to $p = 8.38$ s with such a magnetic field
exceeds the age of the Universe. We supposed that the NS was
born with higher value of magnetic field and that the field
strength has substantially decreased during the evolution. Using
the model of ohmic dissipation of the magnetic field in the NS
crust, we calculated the possible evolution of the NS on the
$B-P$ diagram. The observed rotational period can be obtained at $Q=
0.01-0.05$. However, the evolution of the period depends on the
field dissipation rate and, therefore, on the parameters of the
decay model. Thus, the twofold change in the impurity parameter
$Q$ caused the spin period to change more than by the order of
magnitude at the accretor stage (Fig. 2a, tracks 1 and 2). For
this reason, observations of periods of old INSs may
become an important test for the validity of the model of
evolution of neutron stars.

The decay of the magnetic field may affect the estimate of the
total number of the observed accreting INSs. In particular,
because the formation of a periodic X-ray source at the
propeller stage calls for rather strong magnetic field (Popov,
1994), its decay may reduce the number of sources of this kind.
At the same time, the absence of pulsations for other INS
candidates may suggest that their field has already decayed to
such extent that it cannot funnel the plasma toward the INS
polar caps.

\section*{Acknowledgements}

The authors express their gratitude to F. Haberl for providing them 
information on the X-ray source and to M.E. Prokhorov, V.A.Urpin, and 
V.M. Lipunov for discussion of this work. Special gratitude should be 
expressed to D.G. Yakovlev for carefully reading the manuscript and 
valuable criticism and comments. 
The work of D. Konenkov was supported 
by the Russian Foundation for Basic Research (project 96-02-16905a), 
the work of S. Popov, by the Russian Foundation for Basic Research (project 
95-02-06053a), an INTAS grant (93-3364), and an ISSEP grant (a96-1896). 

\section{References}

\noindent
Blaes, O. and Madau, P., Astrophys. J., 1993, vol. 403, p. 690. 

\noindent
Blaes, O. and Rajagopal, M., Astrophys. J., 1991, vol. 381, p. 210. 

\noindent
B\"ohringer, H., Morfill, G.E. and Zimmermann, H.U., Astrophys. J.,
1987, vol. 313, p. 218. 

\noindent
Friedman, B. and Pandharipande, V.R., Nucl. Phys., 1981, vol. A361, 
p. 502. 

\noindent
Haberl F., Pietsch, W., Motch, C., Buckley, D.A.H., Circ. IAU,
1996, no. 6445. 

\noindent
Illarionov, A.F. and Syunyaev, R.A., Astron. Astrophys., 1975, vol. 39, 
p. 185. 

\noindent
Itoh N., Hayashi H., and Kohyama Y., Astrophys. J., 1993, vol. 418, p. 
405. 

\noindent
Lipunov, V.M., Astrofizika Neitronnykh Zvezd (Astrophysics of 
Neutron Stars), Moscow: Nauka, 1987. 

\noindent
Lipunov, V.M. and Popov, S.B., Astron. Zh., 1995, vol. 72, p. 711. 

\noindent
Ostriker, J.P., Rees, M.J. and Silk, J., Astrophys. Lett., 1970, vol. 6,
p. 179. 

\noindent
Pethick, C.J., Rev. Mod. Phys., 1992, vol. 64, p. 1133. 

\noindent
Popov, S.B., Astron. Tsirk., 1994, no. 1556, p. 1. 

\noindent
Taylor, J., Manchester, R. and Lyne, A., Astrophys. J. Suppl. Ser., 
1993, vol. 88, p. 529. 

\noindent
Treves, A. and Colpi, M., Astron. Astrophys., 1991, vol. 241, p. 107. 

\noindent
Urpin, V., Geppert, U. and Konenkov, D., Astron. Astrophys., 1996, 
vol. 307, p. 807. 

\noindent
Urpin, V. and Konenkov, D., Mon. Not. R. Astron. Soc., 1997, 
vol. 284, p. 741. 

\noindent
Urpin, V.A. and Muslimov, A.G., Astron. Zh., 1992, vol. 69, p. 1028. 

\noindent
Van Riper, K.A., Astrophys. J., 1988, vol. 329, p. 339. 

\noindent
Yakovlev, D.G. and Urpin, V.A., Astron. Zh., 1980, vol. 24, p. 303. 

\noindent
Zdunik, J.L., Haensel, P., Paczynski, B. and Miralda--Escude, J., 
Astrophys. J., 1992, vol. 384, p. 129. 

\end{document}